\documentclass{aa}  
\usepackage{graphicx,upgreek,txfonts,aas_macros,soul}
\usepackage{multirow,lscape}   

\usepackage{hyperref}
\usepackage{pdfpages}

\usepackage{color}
\begin{document} 

\def\xmm {\emph{XMM--Newton}}
\def\cxo {\emph{Chandra}}
\def\swift {\emph{Swift}}
\def\frm {\emph{Fermi}}
\def\igr {\emph{INTEGRAL}}
\def\sax {\emph{BeppoSAX}}
\def\xte {\emph{RXTE}}
\def\rst {\emph{ROSAT}}
\def\asca {\emph{ASCA}}
\def\hst {\emph{HST}}
\def\nst {\emph{NuSTAR}}
\def\gaia {\emph{Gaia}}
\def\srclong {\mbox{CXOU\,J005440.5--374320}}
\def\src {\mbox{J0054}}
\def\flux {\mbox{erg cm$^{-2}$ s$^{-1}$}}
\def\lum {\mbox{erg s$^{-1}$}}
\def\nh {$N_{\rm H}$}
\newcommand{\rev}[1]{ #1}
\newcommand{\revv}[1]{ #1}
\newcommand{\revvv}[1]{{\color{blue} #1}}
\newcommand{\revvvv}[1]{{\color{blue} #1}}
\newcommand{\matt}[1]{\textcolor{red}{#1 - MI}}
\newcommand{\fabio}[1]{\textcolor{violet}{#1 - FP}}


   \title{The restless population of bright X-ray sources of NCG~3621}

\titlerunning{The bright X-ray sources of NGC~3621}
\authorrunning{A. Sacchi et al.}

   \author{A.\,Sacchi,\inst{1} 
   M.\,Imbrogno,\inst{2,3}
   S.\,E.\,Motta,\inst{4,5}
   P.\,Esposito,\inst{6,7} 
   G.\,L.\,Israel,\inst{3}
   N.\,O.\,Pinciroli Vago,\inst{3,8}
   A. De Luca,\inst{7}
   M. Marelli,\inst{7}
   F. Pintore,\inst{9}
   G. A. Rodr\'iguez Castillo,\inst{7}
   R. Salvaterra,\inst{7}
   A. Tiengo\inst{6,7}
    }

   \institute{
Center for Astrophysics $\vert$ Harvard \& Smithsonian, 60 Garden Street, Cambridge, MA 02138, USA\\
e-mail: \href{mailto:andrea.sacchi@cfa.harvard.edu}{andrea.sacchi@cfa.harvard.edu}
\and Dipartimento di Fisica, Università degli Studi di Roma ``Tor Vergata'', via della Ricerca Scientifica 1, I-00133 Roma, Italy
\and INAF--Osservatorio Astronomico di Roma, via Frascati 33, I-00078 Monteporzio Catone, Italy
\and INAF–Osservatorio Astronomico di Brera, via E. Bianchi 46, I-23807 Merate (LC), Italy
\and Department of Physics, Astrophysics, University of Oxford, Denys Wilkinson Building, Keble Road, OX1 3RH Oxford, UK
\and Scuola Universitaria Superiore IUSS Pavia, Palazzo del Broletto, piazza della Vittoria 15, I-27100 Pavia, Italy
\and INAF--Istituto di Astrofisica Spaziale e Fisica Cosmica di Milano, via A. Corti 12, I-20133 Milano, Italy
\and Department of Electronics, Information and Bioengineering,
Politecnico di Milano, via G. Ponzio, 34, I-20133 Milan, Italy
\and INAF--IASF Palermo, via Ugo La Malfa, 153, I-90146, Palermo, Italy
}
              
  \date{Received DD Month YYYY; accepted DD Month YYYY}

  \abstract{We report on the multi-year evolution of the population of X-ray sources in the nuclear region of \revv{NGC}~3621 based on \cxo, \xmm\ and \swift\ observations. Among these, two sources, X1 and X5, after their first detection in 2008, seem to have faded below the detectability threshold, a most interesting fact as X1 is associated with the AGN of the galaxy. Two other sources, X3 and X6 are presented for the first time, the former showing a peculiar short-term variability in the latest available dataset, suggesting an egress from eclipse, hence belonging to the handful of known eclipsing ultra-luminous X-ray sources. One source, X4, previously known for its ``heart-beat'', i.e. a characteristic modulation in its signal with a period of $\approx1$ h, shows a steady behaviour in the latest observation. Finally, the brightest X-ray source in NGC~3621, here labelled X2, shows steady levels of flux across all the available datasets but a change in its spectral shape, reminiscent of the behaviours of Galactic disk-fed X-ray binaries.}
  \maketitle

\section{Introduction}
Ultraluminous X-ray sources (ULXs) are defined as off-nuclear sources with an isotropic luminosity exceeding the $10^{39}$ erg/s threshold (see \citealt{kaaret17,Fabrika2021,King2023} and references therein for recent reviews). The idea behind this definition is to comprise all sources with a luminosity exceeding the Eddington limit for a black hole (BH) of $\approx10\,{\rm M_\odot}$ and to exclude sources powered by accretion onto a supermassive black hole (SMBH), which usually reside in the nuclei of their host galaxies and are hence dubbed active galactic nuclei (AGN).

ULXs are interpreted as binary systems powered by accretion onto a compact object and, given their extreme luminosities, offer the opportunity to study either accretion at or above the Eddington limit for neutron stars (NSs) and stellar mass BHs, or unusually massive BHs, with masses in the range $10^2$--$10^5\,{\rm M_\odot}$, the so-called intermediate-mass black holes (IMBHs).
The \rev{most extreme example} of super-Eddington accretion is probably NGC~5907 ULX, a ULX pulsar containing an accreting NS, whose luminosity exceeds by almost three orders of magnitude its Eddington limit \citep{israel16}.

On the other hand, ESO 243-49 HLX-1, with a peak luminosity as high as $10^{42}$ erg/s is, to date, considered the most convincing candidate IMBH \citep{farrell09}. This source was discovered owing to its luminosity as well as to the fact that it is well far from its host galaxy centre. However, a large fraction of IMBHs are expected to be hosted in the nuclear regions of their host galaxies \citep{chilingarian18}, and even wandering IMBHs are predicted to undergo periods of high luminosity when transiting through galactic nuclei \citep{weller23}. 

\revv{ULXs spectra are generally characterised by two thermal components, usually described with black-bodies or modified accretion disc black-bodies, that dominate the emission above and below $\sim1$~keV. A third, high-energy component, usually modelled with a cut-off power law \citep[e.g.][]{walton18}, is also observed in all the ULX pulsars, suggesting it descends from an accretion column above the NS. In a super-Eddington accretion scenario, the lower energies thermal component is associated with either the emission from an optically thick outflow \citep[e.g.][]{poutanen06} or an outer disc (observed temperatures of $\sim0.2-0.5$~keV), while the higher energy one could be an inner disc emission deprived of winds (temperatures of $\sim1-2$~keV).}

NGC~3621 is a field spiral galaxy in the Hydra constellation, at a distance of 6.7 Mpc \citep{tully13}. Despite being morphologically classified as bulgeless, the nuclear region of NGC~3621, roughly defined as the region within a 2~kpc radius from the galactic centre (corresponding to a projected distance of $1'$), is a most interesting environment. By analysing the optical spectrum and stellar dynamic, \citet{barth09} assigned a Seyfert\,2 classification to the galaxy and inferred an upper limit on the central BH mass of $3\times10^6\,{\rm M_\odot}$. The presence of an AGN in the nucleus of NGC~3621, powered by a particularly light SMBH, was also inferred by the inspection of its infrared emission \citep{satyapal07} and association with an X-ray source \citep{gliozzi09}. The lower limit on the BH mass, inferred from the X-ray luminosity of the source associated with the infrared emission lower limit is about $4\times10^3\,{\rm M_\odot}$, making it possible that the black hole powering the AGN in NGC~3621 is an IMBH.

In the X-ray band, NGC~3621 has been imaged repeatedly. In 2008, the \cxo\ visit (ObsID: 9278, PI: Gliozzi, hereafter dubbed CXO1) that revealed the X-ray counterpart of the AGN, hereafter dubbed X1, also detected two off-nuclear ULXs, with luminosities of about $10^{39}$ erg/s, far from the galaxy's nucleus by about $20''$. \cite{gliozzi09} labelled these two sources B and C, while in this paper they will be addressed as X2 and X5, respectively.

A follow-up observation of NGC~3621, obtained with \xmm\ in 2017 (ObsID: 0795660101, PI: Annuar, hereafter dubbed XMM1), revealed an additional most-interesting source, 4XMM J111816.0–324910, dubbed X4 in this paper. This source was too faint to be noticed in the \cxo\ observation, while in the \xmm\ observation, its luminosity exceeded the $10^{39}$ erg/s threshold, hence earning a classification of transient ULX. Furthermore, in the \xmm\ observation, this source exhibited a peculiar quasi-periodic modulation in its signal, with a period of $\approx1$ h. The X-ray spectral and timing analysis, as well as the search for the optical counterpart of this source, have been presented in  \citet{motta20}.

\begin{figure*}
	\includegraphics[width=\hsize]{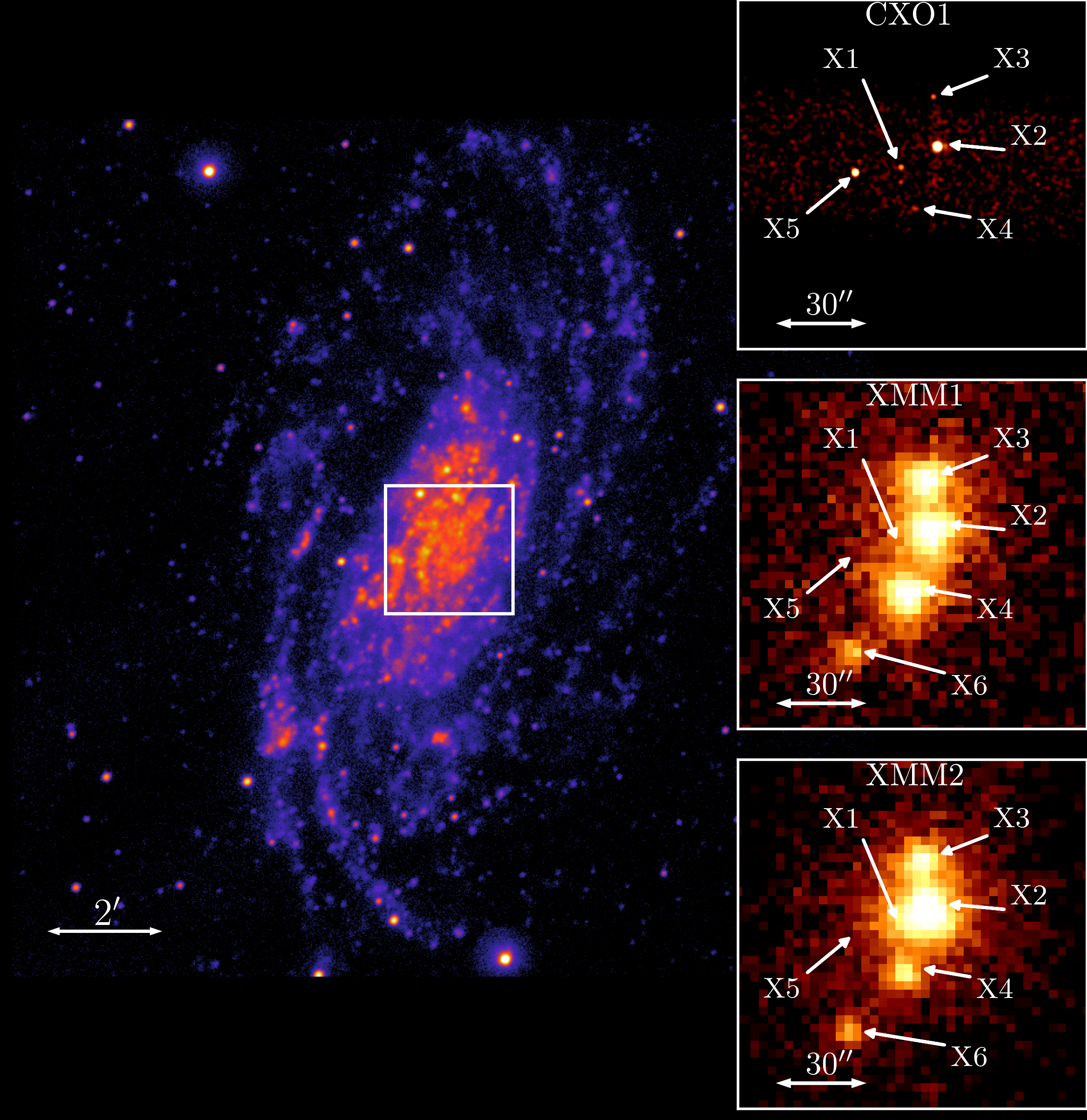}
    \caption{NGC~3621 from NAO-IRAF image in the NUV band. The inset shows the location of the ULXs described in the paper, in the three epochs of the \cxo\ and \xmm\ visits.\label{fig:ngc3621}}
\end{figure*}

The discovery of this last transient ULX prompted extensive \swift/XRT monitoring of NGC~3621, which culminated in a second \xmm\ observation (ObsID: 0884030101, PI: Motta, hereafter XMM2), obtained in May 2023.
Along with X4, two other ULXs (X3 and X6), previously neglected, stand out in both \xmm\ observations, raising the number of X-ray sources surrounding the nuclear region of NGC~3621 to five. This paper is dedicated to the description of the X-ray behaviour of the ULXs residing in the nuclear region of NGC~3621 as well as its central AGN.
Figure \ref{fig:ngc3621} shows NGC~3621 in the near-ultraviolet (NUV) band with three insets to highlight the evolution of the ULX population surrounding its nuclear region. The position and nomenclature of the sources analyzed in this work are summarized in Tab. \ref{tab:names}.

\begin{table*}
\caption{Nomenclature and position for the sources analyzed in this work. \label{tab:names}} 
\centering
\begin{tabular}{cccc}
\hline\hline
label & R.A.$^1$ & DEC.$^1$ & comments \\ 
\hline
X1 & 11:18:16.51 & -32:48:50.4 & the AGN, source A in \citealt{gliozzi09})\\
X2 & 11:18:15.16 & -32:48:40.6 & the brightest source, labelled B in \citealt{gliozzi09}\\
X3 & 11:18:15.30 & -32:48:17.3 & the eclipsing source\\
X4 & 11:18:15.99 & -32:49:10.2 & 4XMM J111816.0–324910, the "heart-beating" source \citep{motta20}\\
X5 & 11:18:18.23 & -32:48:53.0 & source C in \citealt{gliozzi09} \\
X6 & 11:18:17.99 & -32:49:37.3 & the faintest source\\
\hline
\end{tabular}
\tablefoot{$^1$Coordinates are given in ICRS (ep=J2000).} 
\end{table*}

The paper is organized as follows: in Section \ref{sec:data} we describe the dataset upon which the presented analysis is based and the observational properties of each ULX, in Section \ref{sec:disc} we discuss the possible nature of each source and finally in Section \ref{sec:summ} we summarize our results.

\section{Observational properties}\label{sec:data}

\subsection{X-ray dataset}

The results presented in this work rely on all the available X-ray observations of NGC~3621, obtained with \cxo, \xmm, and the \emph{Neil Gehrels Swift Obervatory} and already partially analyzed and described by \citet{gliozzi09} and \citet{motta20}. In addition to these data, the bulk of the analysis is based on the observations of NGC~3621, performed with \swift\ and \xmm\ in the context of a monitoring campaign of X4.

\swift--XRT data were retrieved from the science archive\footnote{\url{https://heasarc.gsfc.nasa.gov/cgi-bin/W3Browse/swift.pl}} and reprocessed using the \texttt{ftool} routine \texttt{xrtpipeline}. The count rates and upper limit for each observation (without separating the individual orbits) were computed with the \texttt{ximage} tool \texttt{sosta} and converted into fluxes adopting the best fitting model of the closest available \xmm\ or \cxo\ observation.

\xmm\ data were downloaded from the science archive\footnote{\url{http://nxsa.esac.esa.int/nxsa-web/\#search}} and reprocessed following the standard procedure with \texttt{sas} v21.0.0 \citep{gabriel04}. The lightcurve of the full observation was inspected and periods of high particle background were excluded from our analysis. For XMM1 we filtered out time intervals during which the background rate was higher than 0.6 and 0.35\,cts/s for PN and MOS, respectively. For observation XMM2, we excluded two high background periods at the beginning and at the end of the observation. Data from the EPIC-MOS cameras were merged using the \texttt{sas} tool \texttt{merge}, once verified that the two cameras provided compatible results. To build spectra and lightcurves, to minimize contamination from nearby sources, counts were extracted from 15"-radii circular regions centred on each source position. 
Background counts were extracted from nearby 30"-radii circular regions, from source-free portions of the same detector chip of each source. Upper limits were obtained using the \texttt{sas} tool \texttt{eupper}. Spectra, spectral redistribution matrices and ancillary response files were generated using the dedicated \texttt{sas} tools.

CXO1\rev{, taken in sub-array mode,} was reprocessed and reduced with the Chandra Interactive Analysis of Observations software package (\texttt{CIAO}, v.4.12; \citealt{fruscione06}) and the \texttt{CALDB} 4.9.0 release of the calibration files. Sources' counts were extracted, using the \texttt{CIAO} tool \texttt{srcflux}, from regions encompassing a point-spread function (PSF) fraction of about 95\% (corresponding to roughly 3"-radius circles). Backgrounds were estimated from source-free annular regions, centred on the source position, of 7 and 15 arcseconds radii. Spectra, spectral redistribution matrices and ancillary response files were generated using the \texttt{CIAO} script \texttt{specextract}.

All reported luminosities are computed assuming the NGC~3621 distance of 6.7 Mpc.

\subsection{Spectral analysis}

Here we report the X-ray spectra and associated spectral analysis of the X-ray bright sources in the nuclear region of NGC~3621. Spectra already published by \citet{gliozzi09} and \citet{motta20} will not be further analyzed. Still, a comprehensive analysis of the multi-epoch spectral evolution of the single sources will be presented in Section \ref{sec:disc}.


The spectra were then fed into the spectral fitting package \texttt{XSPEC} \citep{arnaud96} version 12.12.1. Only counts in the energy range $0.3-10$~keV were considered and, given that all analyzed spectra have sufficient counts, the spectra were rebinned to have at least \rev{20 counts per energy bin and $\chi^2$-statistic was adopted}. All X-ray spectra were corrected for foreground interstellar absorption adopting the model \textsc{TBabs}, with \nh\ fixed to the Galactic value $6.63\times10^{20}$ cm$^{-2}$ \citep{hi4pi16}, using abundances from \citet{wilms00}, with the photoelectric absorption cross-sections from \citet{verner96}, all the reported value of intrinsic absorption are hence to be considered on top of this foreground value. All reported uncertainties correspond to $1\sigma$, all fluxes are intended as observed while luminosities are reported as unabsorbed. \rev{Spectra extracted from CXO1 were analyzed in the $0.5-7$~keV band, but luminosities are all reported in the $0.3-10$~keV band to allow for direct comparison.}

\subsubsection{X1}\label{section:specX1}
The AGN, which is labelled X1, is marginally detected in CXO1, described at length in \citet{gliozzi09}, and not detected in either XMM1 or XMM2, so no further spectral analysis is possible for this source.

\subsubsection{X2}
The X-ray spectrum of X2 from CXO1 is described in \citet{gliozzi09}, where it is reproduced by an absorbed power law with \nh$=(3.4\pm0.6)\times10^{21}$ cm$^{-2}$ and $\Gamma=2.7\pm0.3$. 



The X-ray spectrum of X2 can be reproduced by an absorbed power law in XMM1 as well. The best fitting parameters are \nh$=(1.1\pm0.1)\times10^{21}$ cm$^{-2}$ and $\Gamma=1.87\pm0.03$, corresponding to a statistic of \rev{$\chi^2/\nu=241.54/188\approx1.27$ (where $\chi^2$ is the value of the $\chi^2$-statistic and $\nu$ the number of degrees of freedom)}. 



In XMM2 the X-ray spectrum of X2 cannot be reproduced satisfactorily with a power law \rev{($\chi^2/\nu=1078.34/263\approx4.1$)}, nor by a combination of a power law and other models. Instead, it can be well reproduced by an absorbed multi-colour disc model, \textsc{diskbb} in \texttt{Xspec}, with $\chi^2/\nu=287.63/263\approx1.09$. \revv{Although not specifically requested by the data, we performed a further spectral fit by adding a second thermal component, also modelled with a \textsc{diskbb}, to be in line with the most recent findings on ULX spectral properties. We note that the fit is marginally improved ($\chi^2/\nu=280.71/261\approx1.07$).} 


\revv{At this point, we tried to fit the data from CXO1 and XMM1 with the same absorbed two-thermal component model. We obtained acceptable fits for both epochs, with marginally better $\chi^2$ with respect to the single power-law model.}

\begin{figure*}
	\includegraphics[width=\hsize]{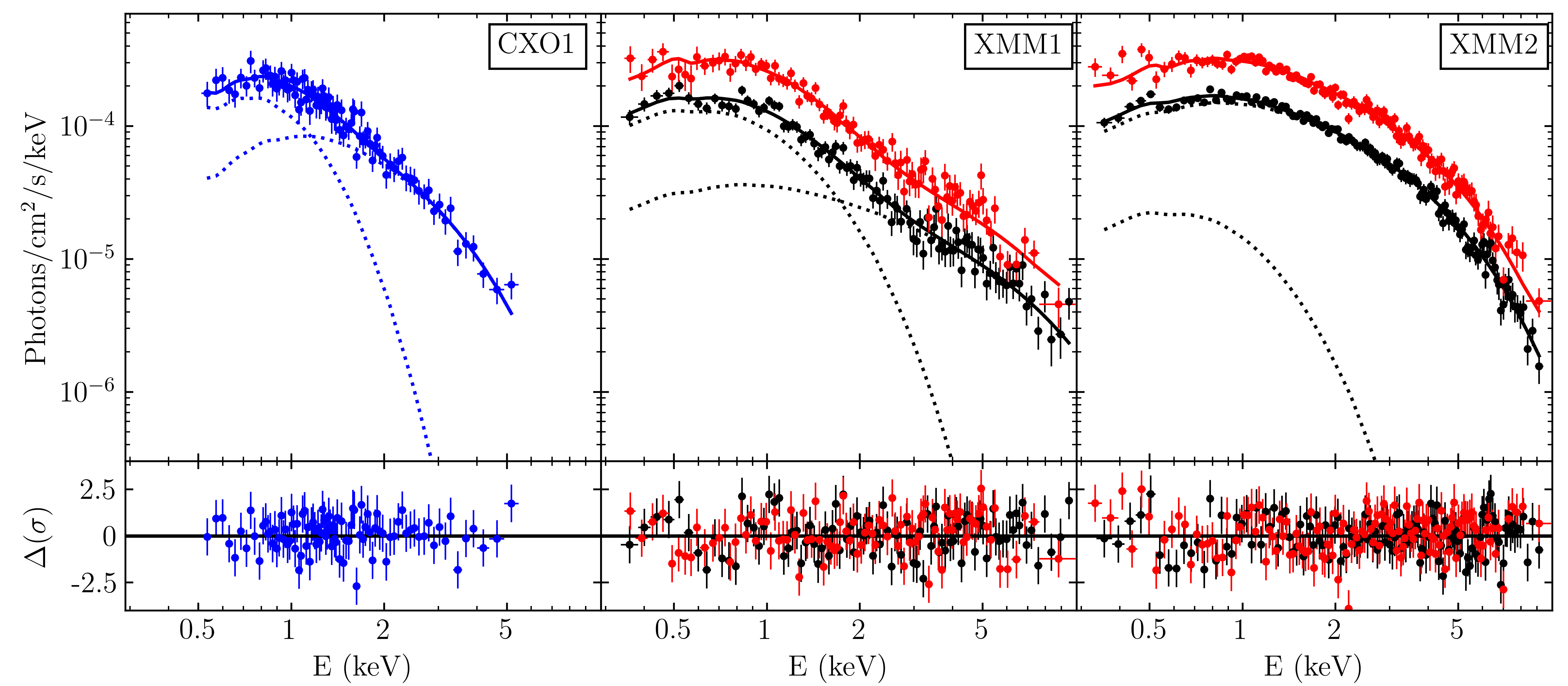}
    \caption{X-ray spectrum (upper panels) and residuals (lower panels) of the source labelled X2 in the three different epochs. In blue \cxo\ data, in black \xmm/pn and red merged \xmm/MOS data. The solid lines show the best-fitting models described in the text. \revv{The individual components contributing to the model are shown in dotted lines (for XMM1 and XMM2 the two components are shown only for the pn dataset for clarity)}.
    \label{fig:x2_spec}}
\end{figure*}

The best fitting parameters for the two-thermal-component model are reported in Tab. \ref{tab:xspec} and the spectrum in the different epochs is shown in Fig. \ref{fig:x2_spec}.

\subsubsection{X3}
In CXO1, the source labelled X3 was marginally detected. The low number of counts prevents us from performing the spectral analysis but by converting the count rate into flux with the best-fitting model described below, we obtain $\log F=-13.8\pm0.1$ erg/s/cm$^2$, corresponding to a luminosity of $\log L=38.3\pm0.1$ erg/s.


In XMM1 and XMM2, the X-ray spectrum of X3 can be well reproduced by an absorbed power law \rev{($\chi^2/\nu=311.67/273\approx1.14$)}. The best-fit parameters of the model are \nh$=(0.38\pm0.05)\times10^{21}$ cm$^{-2}$ and $\Gamma=2.18\pm0.03$. Data from the two epochs were fitted simultaneously and the goodness of the fit does not improve freeing either of the two parameters in the two different epochs.

\begin{figure}
	\includegraphics[width=\hsize]{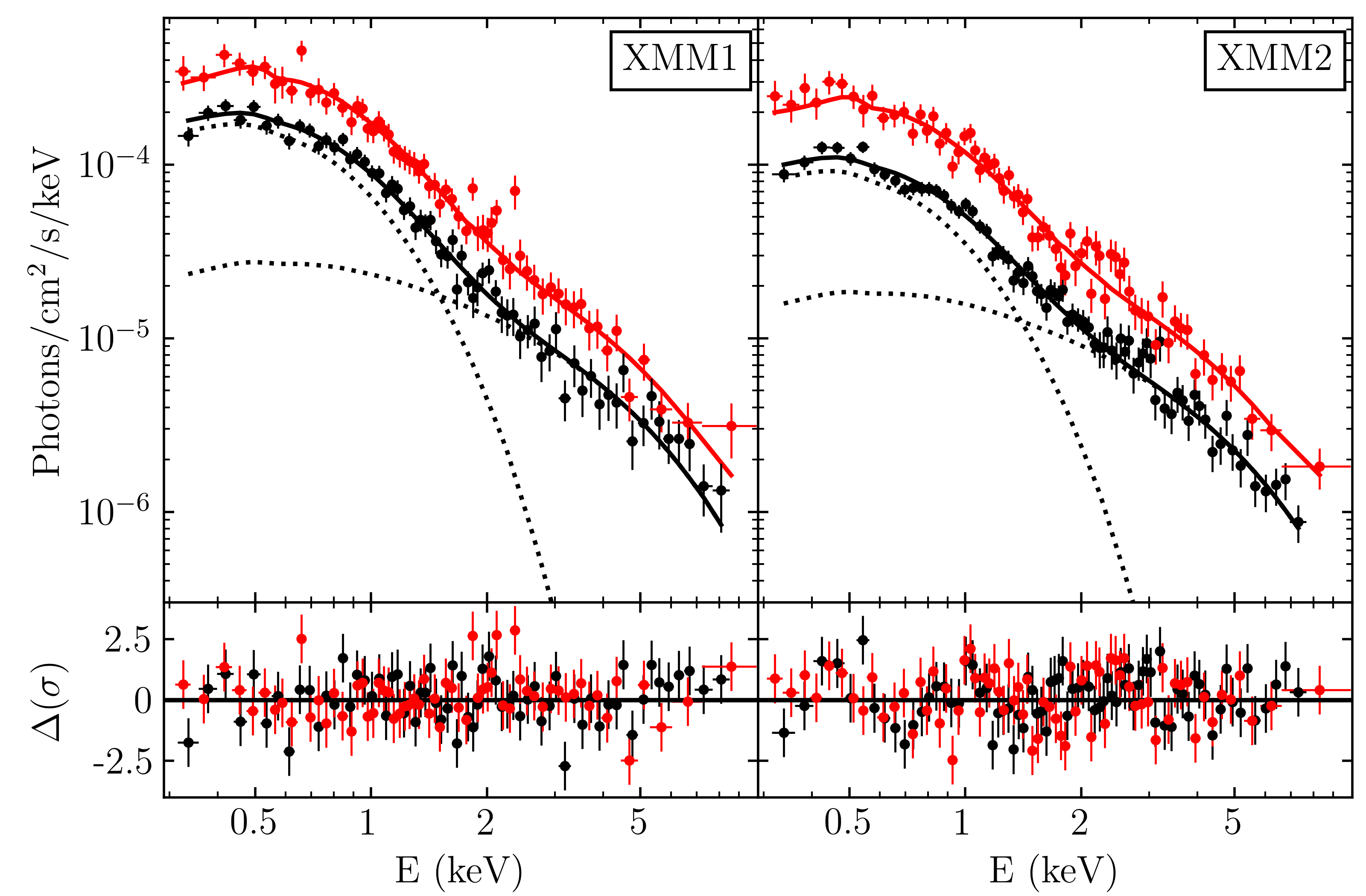}
    \caption{X-ray spectrum (upper panels) and residuals (lower panels) of the source labelled X3 in XMM1 and XMM2 (left and right panels respectively). In black \xmm/pn and red merged \xmm/MOS data. The solid lines show the best-fitting models described in the text, \revv{the dotted lines indicate the two thermal components for the sole pn data.}\label{fig:x3_spec}}
\end{figure}

\revv{Following the strategy adopted for X2, we test whether the fit quality can be improved by adopting a two thermal component model (\textsc{diskbb}+\textsc{diskbb}). We indeed obtain a better statistic ($\chi^2/\nu=273.43/270\approx1.01$), requiring no intrinsic absorption.} The best-fitting parameters are reported in Tab. \ref{tab:xspec} and the spectrum in the different epochs is shown in Fig. \ref{fig:x3_spec}.


\subsubsection{X4}\label{sec:specX4}
The X-ray emission of X4 up until late 2017 has been fully presented and discussed in a dedicated publication by \citet{motta20}.

\begin{figure}
	\includegraphics[width=\hsize]{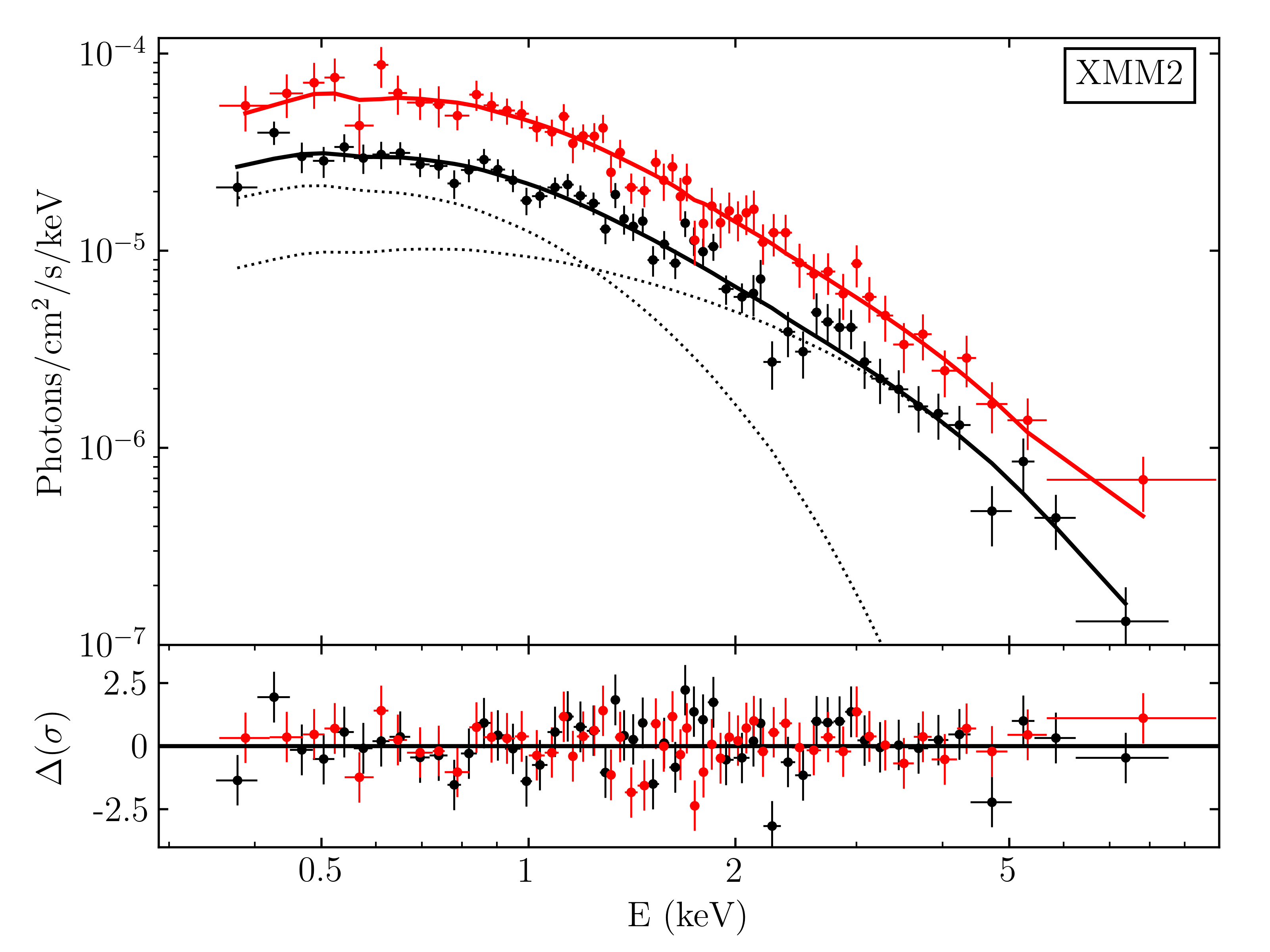}
    \caption{X-ray spectrum (upper panels) and residuals (lower panels) of the source labelled X4 in XMM2. In black EPIC/pn and red merged EPIC/MOS data. The solid lines show the best-fitting models. \rev{The dotted lines indicate the two components for the pn data}.\label{fig:x4_spec}}
\end{figure}

In XMM2, the X-ray spectrum of X4, shown in Fig. \ref{fig:x4_spec}, can be well reproduced by \revv{a two thermal components model (\textsc{diskbb}+\textsc{diskbb}), with statistic $\chi^2/\nu=94.55/103\approx0.92$. The best-fitting parameters of the model are reported in Tab. \ref{tab:xspec}. To allow for direct comparison with the model adopted for this source in \citet{motta20}, we also fit its spectrum with a \textsc{diskbb}+\textsc{bbody} model, obtaining a good fit  ($\chi^2/\nu=108.48/103\approx1.05$) with a blackbody temperature for the soft component of $0.26\pm0.01$~keV.}

 


\subsubsection{X5}\label{sec:specX5}
The X-ray spectrum of X5 as it appeared in CXO1 is described in detail in \citet{gliozzi09} and it is reproduced by an absorbed power law with \nh$=5.1^{+2.5}_{-1.9}\times10^{21}$ cm$^{-2}$  and $\Gamma=1.6\pm0.4$. Its flux in the 0.3--10 keV band is $\log F=-12.74\pm0.04$ erg/s/cm$^2$ and its luminosity $\log L=39.36\pm0.04$ erg/s.

The source is not detected in any of the two subsequent \xmm\ observations, hence no further spectral analysis is possible.

\subsubsection{X6}
X6 position fell off the field of view in CXO1, and in the gap between two detector's chips of the EPIC/pn camera in XMM1, hence we focus only on the EPIC/MOS data of XMM1 and XMM2 and EPIC/pn data of XMM2.


Data were analyzed simultaneously, with all parameters linked together safe for the normalization, and the X-ray spectrum of X6 can be acceptably reproduced by a simple power-law model, with no additional absorption. We obtained \rev{$\chi^2/\nu=102.92/92\approx1.12$}. The best-fitting slope of the power law is $\Gamma=1.72\pm0.03$. 

\begin{figure}
	\includegraphics[width=\hsize]{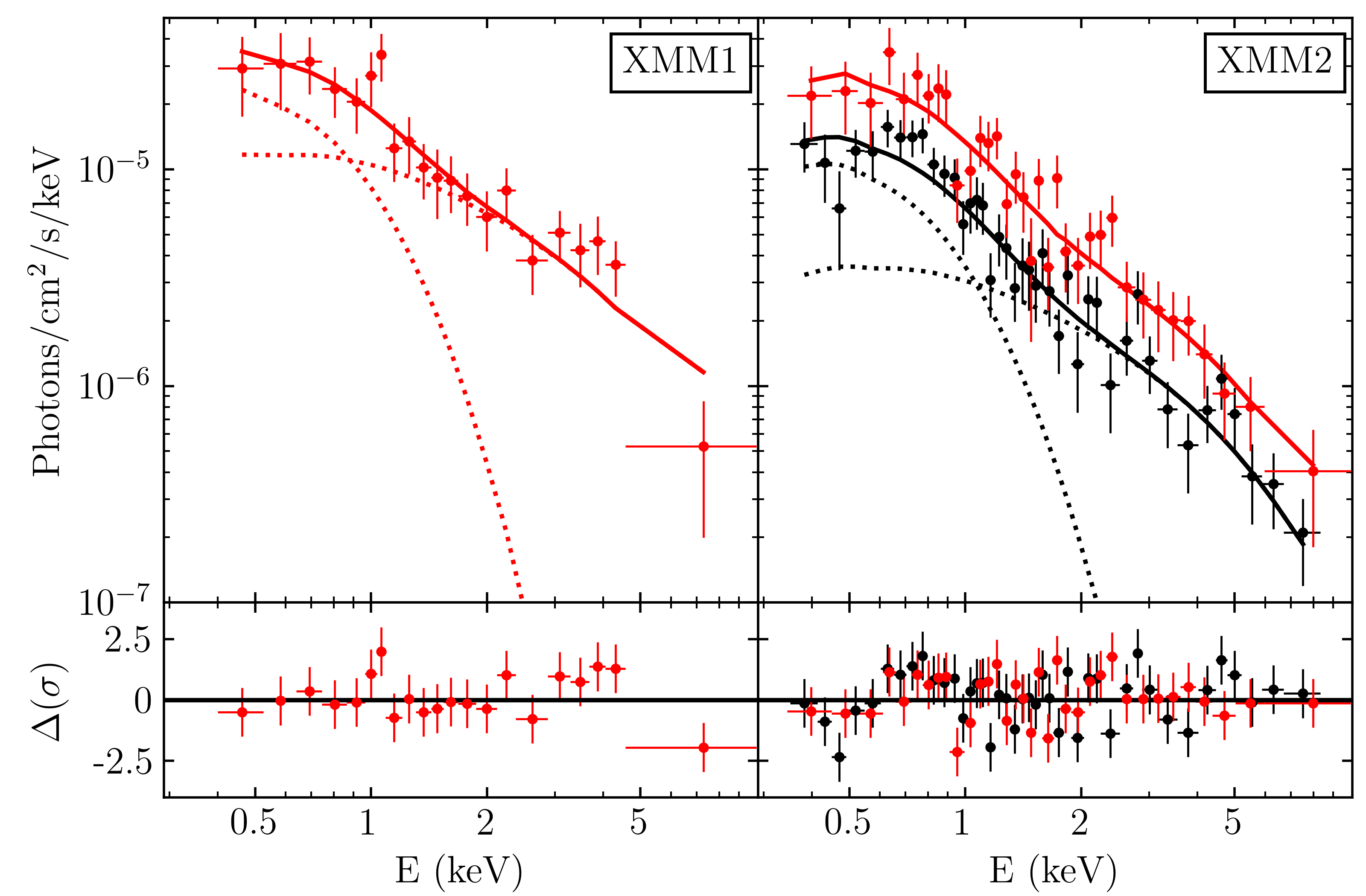}
    \caption{X-ray spectrum (upper panels) and residuals (lower panels) of the source labelled X6 in XMM1 and XMM2 (left and right panels respectively). In black \xmm/pn and red merged \xmm/MOS data. The solid lines show the best-fitting models described in the text, \revv{ the dotted lines indicate the two thermal components}.\label{fig:x6_spec}}
\end{figure}

\revv{In this case too, although the fit is already acceptable with a single power law model, we fit a two thermal component model to the source, which resulted in a better description of the data ($\chi^2/\nu=91.83/90\approx1.02$). Best-fitting parameters are shown in Tab. \ref{tab:xspec} and the spectrum in the different epochs is shown in Fig. \ref{fig:x6_spec}.}


\begin{table*}
\caption{\rev{Summary of the X-ray spectral properties of the analyzed sources. \label{tab:xspec}} }
\centering
\begin{tabular}{c|c|cccccc|c}
\hline\hline
source & observation & $\chi^2/\nu$ & $N_\textup{H}^1$ & $kT_\textup{soft}^2$ & $L_\textup{X,soft}^3$ & $kT_\textup{hard}^2$ & $L_\textup{X,hard}^3$ & $L_\textup{X,tot}^4$ \\ 
\hline
\multirow{3}{*}{X2} & CXO1 & 0.75 & $2.4\pm0.8$ & $0.25\pm0.06$ & $3.2\pm0.8$ & $1.0\pm0.1$ & $3.6\pm0.4$ & $6.5\pm0.3$ \\
                    & XMM1 & 1.13 & $0.6\pm0.1$ & $0.43\pm0.03$ & $1.70\pm0.08$ & $2.8\pm0.3$ & $3.24\pm0.07$ & $4.97\pm0.09$\\ 
                    & XMM2 & 1.07 & $0.8\pm0.1$ & $0.37\pm0.11$ & $0.2\pm0.1$ & $1.57\pm0.02$ & $7.9\pm0.1$ & $8.13\pm0.07$\\
\hline
\multirow{2}{*}{X3} & XMM1 & \multirow{2}{*}{1.01} & \multirow{2}{*}{$<0.1$} & \multirow{2}{*}{$0.31\pm0.01$} & $1.07\pm0.01$ & \multirow{2}{*}{$1.9\pm0.1$} & $1.32\pm0.06$ & $6.6\pm0.1$\\ 
                    & XMM2 & & & & $0.57\pm0.01$ & & $0.89\pm0.02$ & $3.83\pm0.06$\\
\hline
X4 & XMM2 & 0.92 & $0.4\pm0.2$ & $0.39\pm0.07$ & $0.21\pm0.04$ & $1.3\pm0.1$ & $0.38\pm0.04$ & $0.59\pm0.01$\\ 
\hline
\multirow{2}{*}{X6} & XMM1 & \multirow{2}{*}{1.02} & \multirow{2}{*}{<5.9} & \multirow{2}{*}{$0.28\pm0.03$} & $0.08\pm0.02$ & \multirow{2}{*}{$2.1\pm0.3$} & $0.31\pm0.04$ & $0.4\pm0.1$\\ 
                    & XMM2 & & & & $0.063\pm0.008$ & & $0.19\pm0.02$ & $0.3\pm0.1$\\
\hline
\end{tabular}
\tablefoot{$^1$Intrinsic absorption in units of $10^{21}$~cm$^{-2}$, \revv{the upper limits are reported at a $3\sigma$ level. $^2$Temperature of the soft and hard thermal components in keV. $^3$X-ray luminosity of the soft and hard thermal components in the 0.3-10~keV band in units of $10^{39}$~erg/s. $^4$Total X-ray luminosity in the 0.3-10~keV band in units of $10^{39}$~erg/s}.} 
\end{table*}

\subsection{Long-term lightcurve}

\begin{figure}
	\includegraphics[width=\hsize]{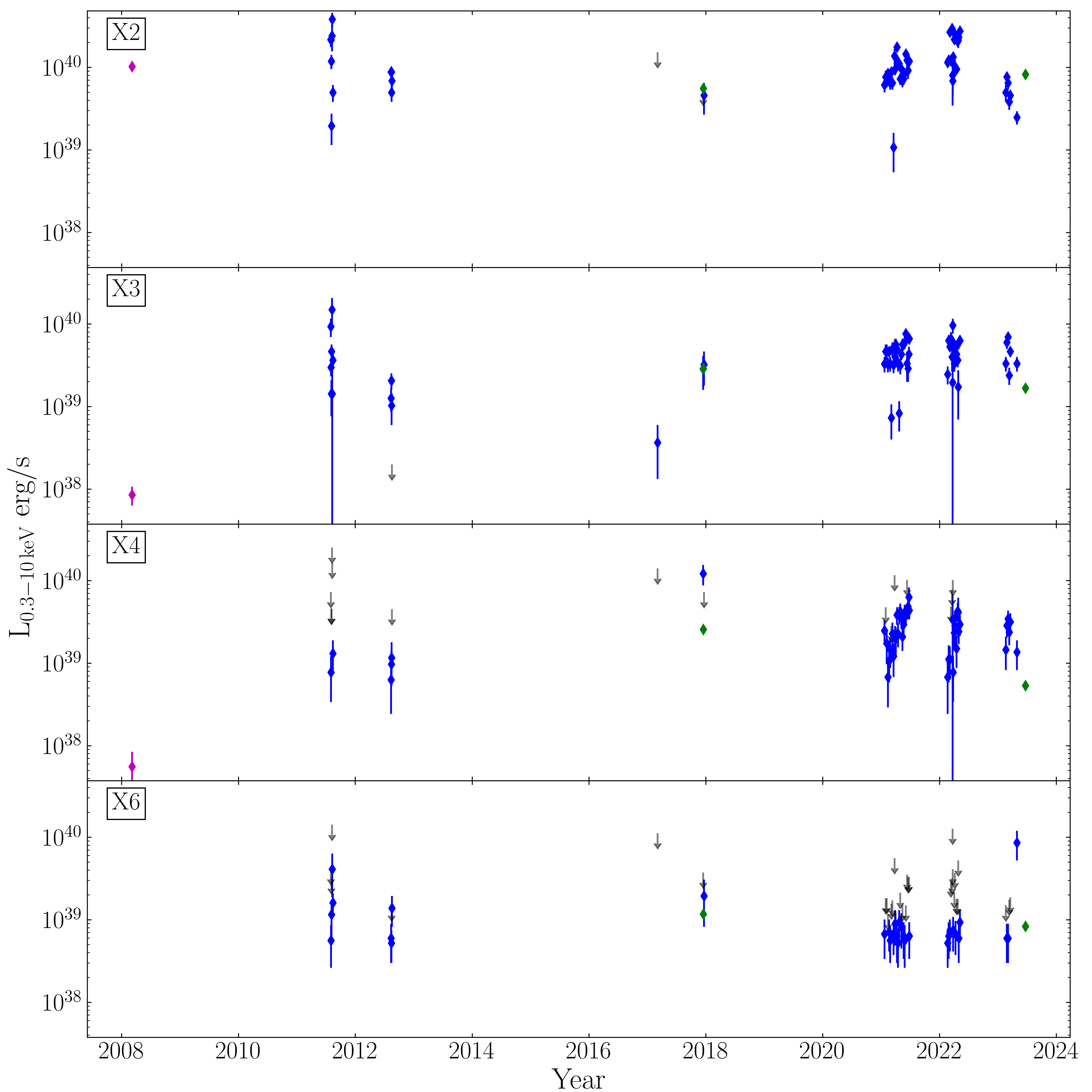}
    \caption{The long-term X-ray lightcurves of the source X1 to X6 (from top to bottom). Magenta, green and blue points indicate \cxo, \xmm\ and \swift/XRT data, respectively. Down-pointing arrows indicate $3\sigma$ upper limits. \swift/XRT upper limits are in grey. \xmm/pn and \xmm/MOS upper limits are shown by two different arrows. The count rates on which this figure is built are reported in Appendix B. \label{fig:lc}}
\end{figure}

The X-ray lightcurves of \revv{four} of the six sources described here are shown in Fig. \ref{fig:lc}, which reports the luminosity in the $0.3-10$~keV band as a function of the epoch of observation. The luminosities were obtained by converting the count rates listed in Tab.~B.1 adopting the best-fitting models described above.

All \revv{presented} sources show significant variability, often by more than one order of magnitude. 

\rev{X1 and X5 are not reported, as these two sources fall within the PSF profiles of the much brighter sources X2 and X4, contaminating them in all \swift/XRT observations. Both X1 and X5 are not detected in any of the two \xmm\ observations}.

\subsection{Timing analysis}

In this section, we report the timing analysis of the two \xmm\ observations, focusing on X2, X3, X4 and X6, as X1 and X5 were not detected (see previous section). We used the \textsc{SAS} tool \texttt{evselect} to extract source events in the 0.3--10\,keV band. The light curves and the power density spectra (PDSs) are computed through the \textsc{XRONOS} tasks \texttt{lcurve} and \texttt{powspec}, respectively. The light curves shown in this section are all background subtracted and the bin time has been adjusted to ensure there are at least 10 counts per bin. Unless otherwise stated, for the analyzed sources this requirement corresponds to a bin time of 300\,s. A search for coherent periodicity correcting for both a (circular) orbital motion and a first derivative $\dot{P}$ of the (possible) spin signal through Particle Swarm Optimization, an evolutionary algorithm (Pinciroli Vago et al., in prep.), and accelerated search techniques \citep[we followed][]{rodriguez2020} gave negative results. In this section, we report the 3$\sigma$ upper limits on the pulsed fraction of a signal in the PDSs, computed following \cite{israel96} from the Nyquist frequency $\nu_\mathrm{Ny}$ down to 1\,mHz. The PDSs were computed with the maximum combined time resolution, which in the case of PN+MOS data corresponds to 2.7\,s \rev{($\nu_\mathrm{Ny}\simeq0.18$\,Hz)}. For each PDS, we also checked the geometrically rebinned version with a factor 1.16 and 1.08 (each bin is 16\% and 8\% larger, respectively, than the previous one) for broad-band features associated with incoherent variability. We found that every PDS was dominated by white noise. 

Unless otherwise stated, the errors we report in this section correspond to 1$\sigma$ (68.3\%) confidence ranges. \rev{For ease of comparison, we summarise the results of our timing analysis \revv{(in particular the upper limits on the pulsed fraction and the rms fractional variation)} in Tab.~\ref{tab:timing}.}

\begin{table}
\caption{\rev{Summary of the X-ray pulsed fraction and rms fractional variation upper limits derived by our analysis. \label{tab:timing}} }
\centering
\begin{tabular}{cccccc}
\hline\hline
source & observation & PF$^1$ & rms$^2$ \\
\hline
\multirow{2}{*}{X2} & XMM1 & $\lesssim22\%$ & $\lesssim0.078$\\ 
                    & XMM2 & $\lesssim13\%$ & $\lesssim0.044$\\
\hline
\multirow{2}{*}{X3} & XMM1 & $\lesssim30\%$ & $\lesssim0.16$\\ 
                    & XMM2$^3$ & $\lesssim28\%$ & -- \\
\hline
X4 & XMM2 & $\lesssim35\%$ & $\lesssim0.28$ \\ 
\hline
\multirow{2}{*}{X6$^4$} & XMM1 & $\lesssim100\%$ & $\lesssim0.57$\\ 
                    & XMM2 & $\lesssim90\%$ & $\lesssim0.45$ \\
\hline
\end{tabular}
\tablefoot{$^1$ 3$\sigma$ upper limit on the pulsed fraction of a signal in the 1\,mHz--0.18\,Hz range. $^2$3$\sigma$ upper limit on the rms fractional variation in the 10\,\textmu Hz--3\,mHz range. $^3$Given the presence of the eclipse, for this observation we did not derived the rms fractional variation. $^4$ For the light curves of this source we considered a bin time of 2000\,s, given the lower count statistics. Correspondingly, the rms fractional variation is computed in the 10\,\textmu Hz--0.5\,mHz range.} 
\end{table}

\subsubsection{X2}

The timing analysis of X2 data from 2017 to 2023 reveals no significant evolution or distinctive features. The top and bottom panels in the left column of Fig.~C.1 show the 0.3--10\,keV band PDS and the light curve in XMM1, respectively. The top and bottom panels in the right column of Fig.~C.1 show the same plots in the same order in XMM2.

\subsubsection{X3}

X3, together with X4, shows the most interesting evolution from a timing point of view. The 0.3--10\,keV band PDS and the light curve of XMM1 data are shown, respectively, in the \rev{left} and \rev{right} panel of Fig.~C.2. No significant feature is detected. 

\begin{figure}
        \includegraphics[width=\columnwidth]{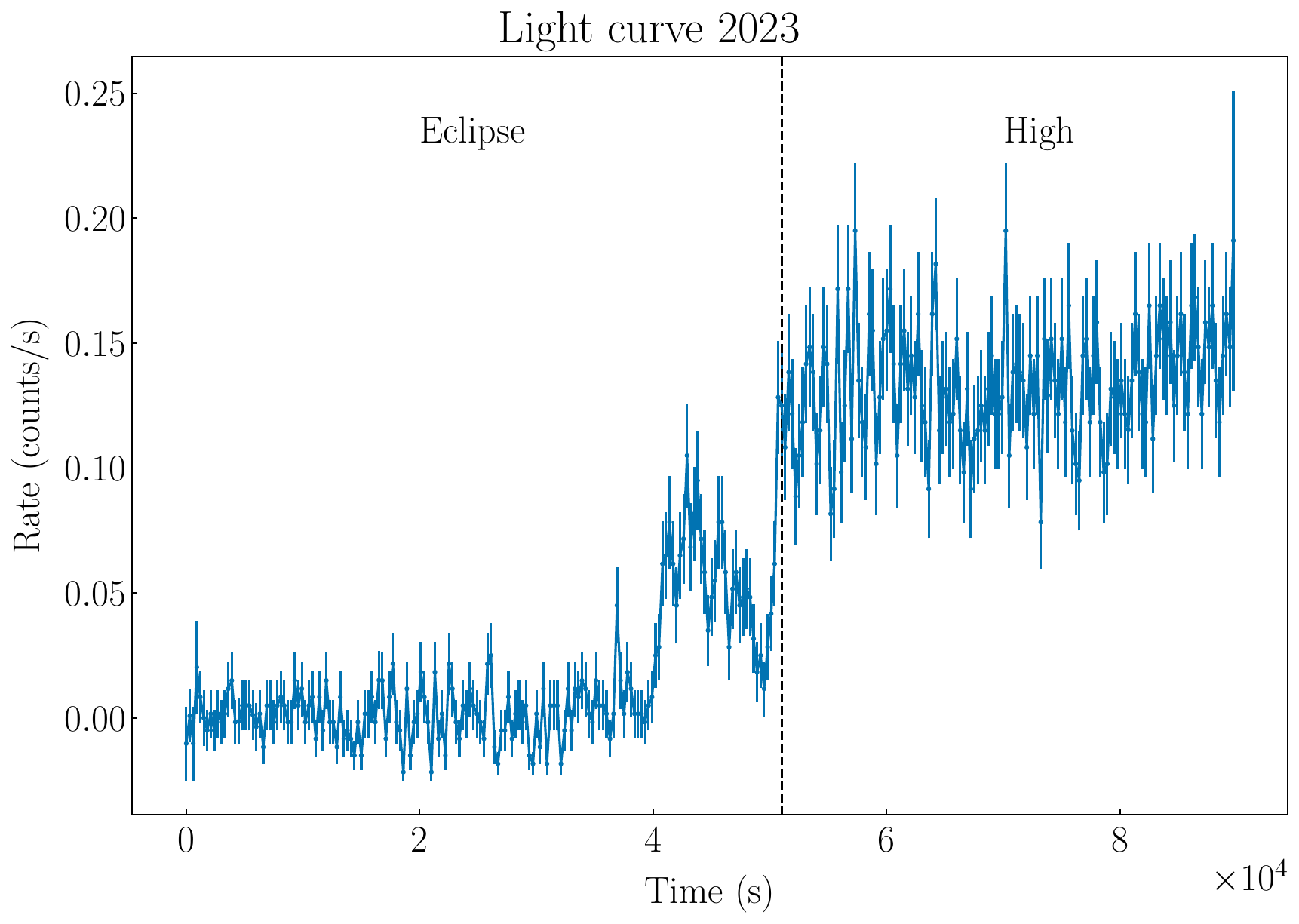}
    \caption{\rev{X3 PN+MOS light curve in the 0.3--10\,keV band during XMM2. The black dashed line at 51\,ks divides the XMM2 light curve into the two regions we considered for our timing analysis. The bin time of the background-subtracted light curve is 300\,s.}\label{fig:x3eclipse}}
\end{figure}

On the other hand, the 0.3--10\,keV band light curve from XMM2 (Fig.~\ref{fig:x3eclipse}) shows what seems to be the egress of an eclipse. During the first 40\,ks of the observation the mean count rate is consistent with a zero flux level. From this, we can derive a lower limit on the eclipse duration of approximately 40\,ks. The count rate then rapidly rises for approximately 3\,ks, only for it to decay back to lower flux rates. After another 1\,ks the count rate increases to a steady level of $(1.30\pm0.02)\times10^{-1}$\,cts/s. 

\begin{figure}
    \centering
    \includegraphics[width=\columnwidth]{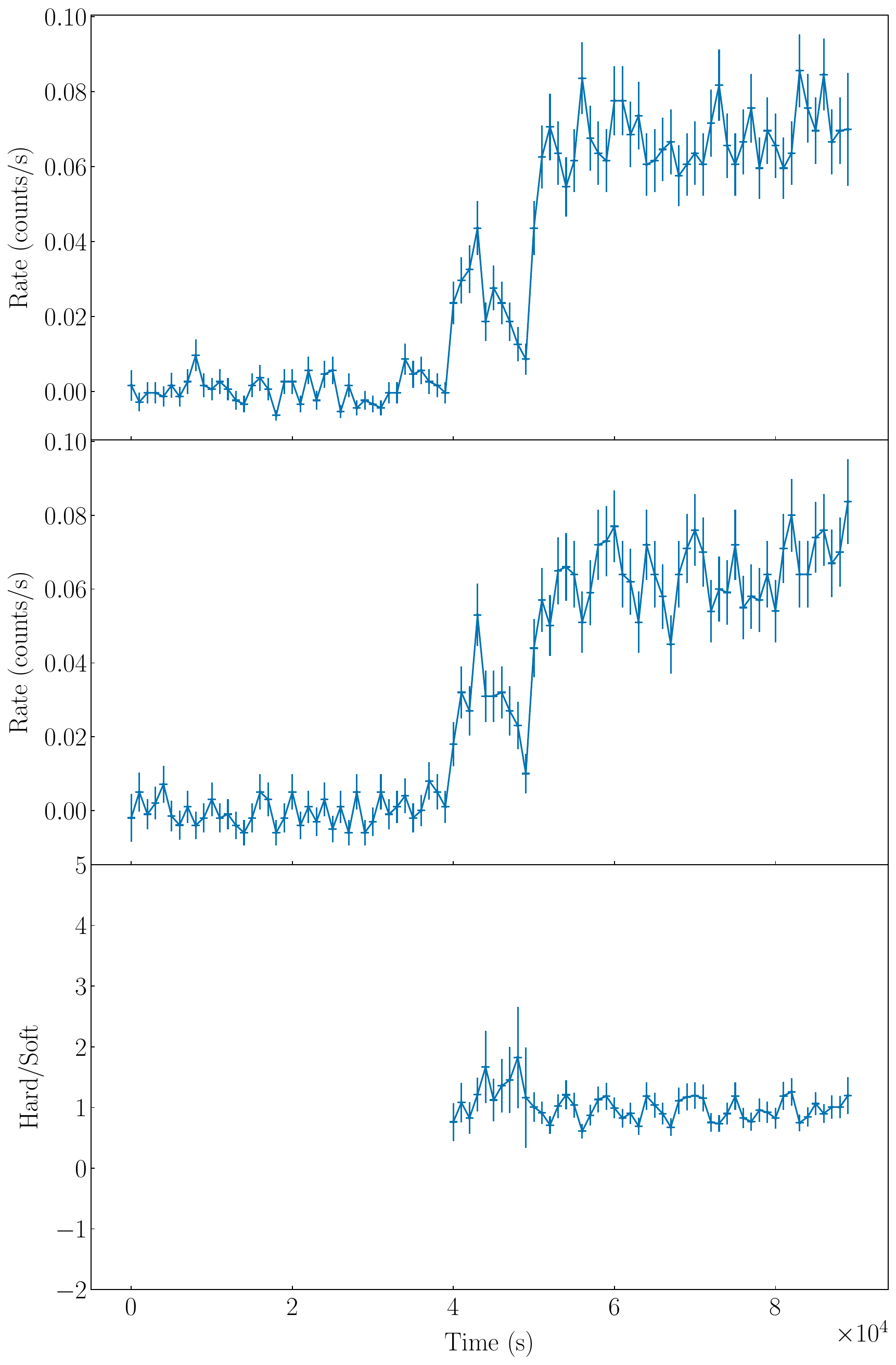}
    \caption{Top panel: X3 PN+MOS soft (0.3--1\,keV) light curve. Middle panel: X3 PN+MOS hard (1--10\,keV) light curve. Bottom panel: hardness ratio of the hard count rate over the soft count rate. 
    We do not show the points in the first 40\,ks since the hardness ratio in this range is dominated by noise. 
    The bin time of the background-subtracted light curves is 1000\,s. 
    }
    \label{fig:x3hr}
\end{figure}

    In Fig.~\ref{fig:x3hr} we report the background-subtracted light curves in the soft (0.3--1\,keV) and hard (1--10\,keV) band, together with the corresponding hardness ratio (hard/soft). We chose these two bands to have a similar number of photons in both light curves. In this case, we adopted a bin time of 1000\,s. 
    We do not show the hardness ratio in the first 40\,ks of the observations since in this time interval it is dominated by noise \rev{due to the low number of collected counts}. No clear evolution is visible in the hardness ratio, which can be fitted with a constant. Although a slight rise is visible at $\simeq40-50$\,ks, its significance is low ($<3\sigma$).

Given the low count rate during the first 51\,ks of observations ('Eclipse' interval) and to avoid artefacts introduced by the eclipse, we computed the PDS (Fig.~C.3) only after the first 51\,ks ('High' interval). We detect no particular feature in the PDS. No coherent signals are detected in both PDSs.

\subsubsection{X4}

Timing analysis of X4, up until XMM1 included, has been fully presented and discussed by \cite{motta20}. In this section, we focus on XMM2.

\begin{figure}
	\includegraphics[width=\columnwidth]{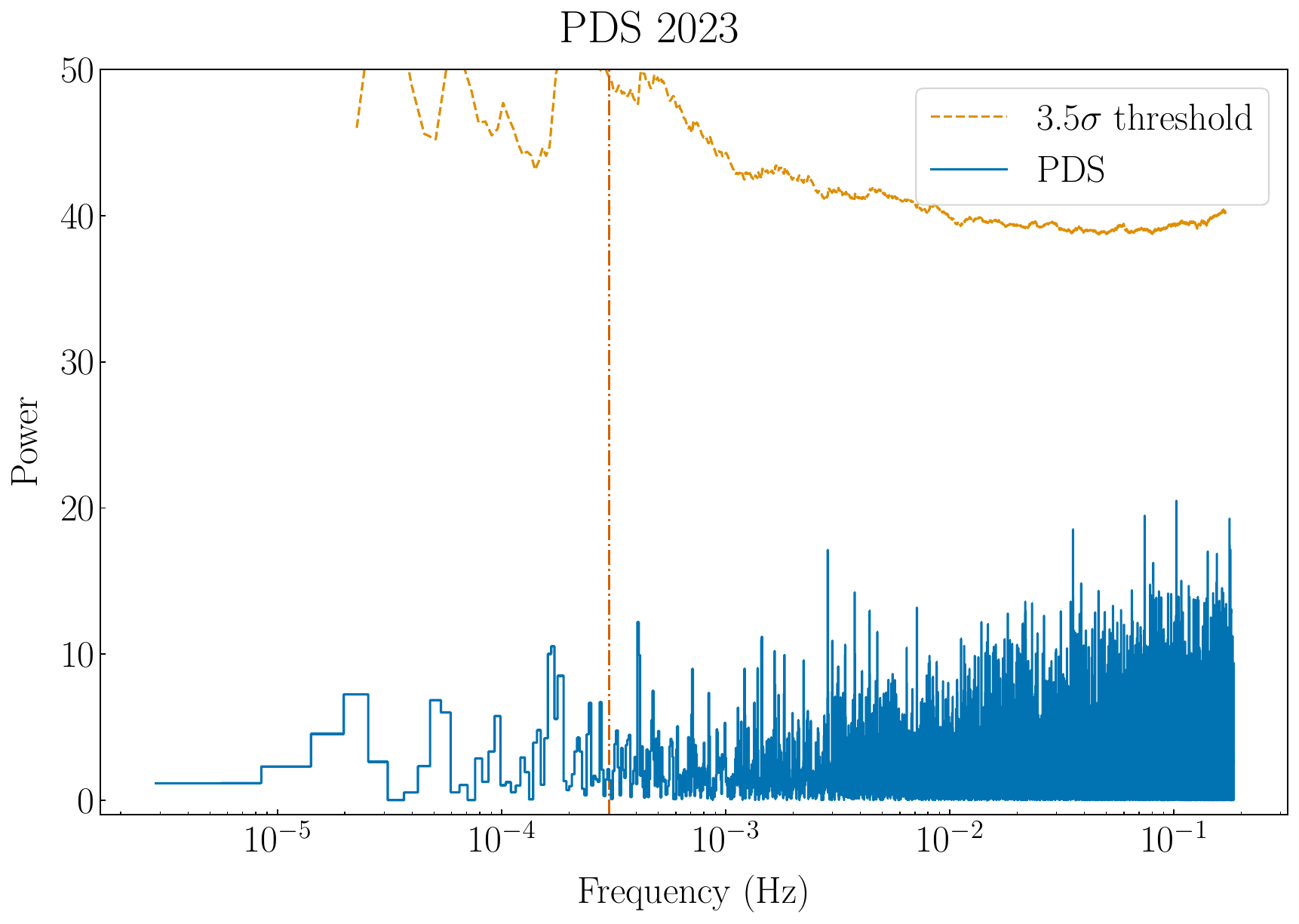}
    \caption{\rev{X4 PN+MOS power density spectrum 
    in the 0.3--10\,keV band during XMM2. The dash-dotted line at $\simeq0.3$\,mHz shows the frequency of the main peak detected in the PDS in XMM1. The yellow dashed line in the original PDS shows the 3.5$\sigma$ detection threshold.}
    \label{fig:x4pds}}
\end{figure}

In XMM1, a striking feature of this source was the presence of a repeating pattern similar to the so-called "heart-beat" of GRS\,1915+105 \citep{Belloni00}. This feature, however, was absent during the 2023 \xmm\ observation, as the 0.3--10\,keV band PN+MOS light curve shows in Fig.~C.4. Fig.~\ref{fig:x4pds} shows the PDS of 2023 data. For comparison, the dash-dotted line in the PDS shows the frequency of the main peak detected in the PDS in XMM1 data. No significant features can be recognised in the XMM2 PDS. 

\subsubsection{X6}

For XMM1, we considered only data coming from the two MOS cameras (see previous Section). The 0.3--10\,keV band light curve (bin time of 2000\,s) and PDS of XMM2 data are shown in the bottom and top panel in the left column of Fig.~C.5.

The same plots derived from 2023 data are shown in the two panels in the right column of Fig.~C.5. In both observations the source was faint, as can be seen from the light curves, in which time bins consistent with zero count rate can be recognised. The source shows no significant evolution between the two observations. 

\subsection{Optical counterparts}
\begin{figure*}
 	\includegraphics[width=\textwidth]{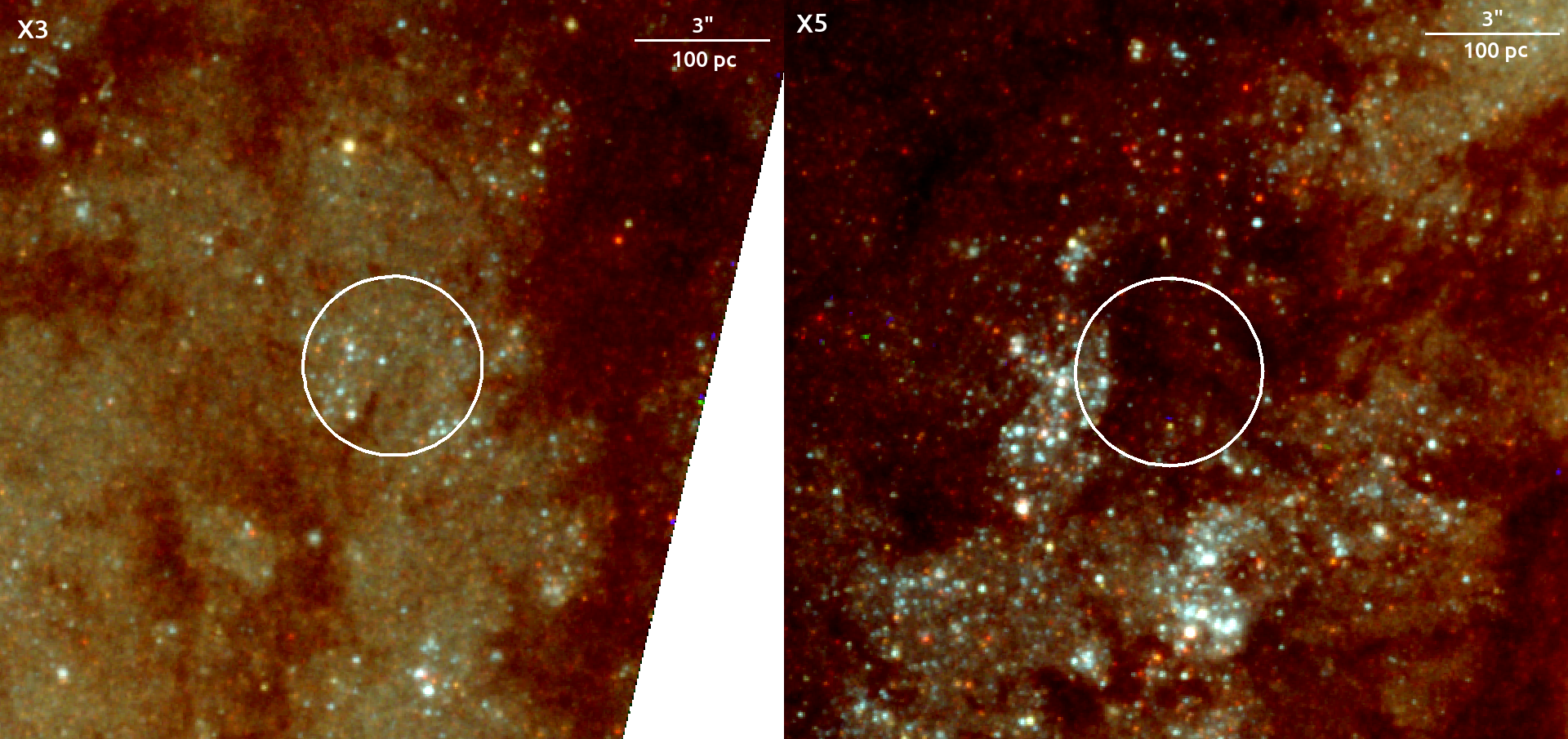}
     \caption{Three-color composite images of the location of the sources labelled X3 and X5 (on the left and right respectively) using the \hst\ F814W, F555W, and F435W filters. The white circles show the PSF size (95\% uncertainty corresponding to $r\approx2''$) of \cxo. \label{fig:oc}}
 \end{figure*}

To attempt an identification of the optical counterpart of the X-ray sources described above, we retrieved the calibrated \hst\ ACS/WFC images of NGC~3621 in three bands: F435W (B), F555W (V), and F814W (I). Unfortunately, \revv{no bright source (e.g. a star or background AGN) is present in both the CXO1 and \hst\ dataset to align the images} and no obvious optically-bright counterpart is present in the error circles of the X-ray detections as illustrated in Fig. \ref{fig:oc}, which shows the RGB HST image with superposed the \cxo\ error circles for the sources labelled X3 and X5.

\section{Discussion}\label{sec:disc}

NGC~3621 hosts in its nuclear region a most-interesting population of X-ray sources which, across three long X-ray observations performed with \cxo\ and \xmm, and several short \swift\ pointings, revealed a plethora of different behaviours. 
While four sources have been already studied to some extent and a part of the pertinent X-ray data are publicly available \citep{gliozzi09,motta20}, in this work, we present a comprehensive description of the ULX population of NGC~3621. 
We base the bulk of this analysis on proprietary data obtained in the monitoring campaign of one of the presented sources (P.I. Motta). 

In this Section, we discuss, for each source, its possible nature in the light of its X-ray behaviour.

\subsection{X1: the AGN}
The presence of an AGN in the central region of NGC~3621 has been first reported owing to {\em Spitzer} detection of emission lines, incompatible with a burst of star formation \citep{satyapal07}. Archival optical images obtained with \hst\ allowed to indicate $3\times10^6\,{\rm M_\odot}$ as the upper limit to the mass of the black hole powering the AGN \citep{barth09}.

The source labelled here as X1 is marginally detected in the 2009 \cxo\ observation, and it is spatially coincident with the mid-infrared emission associated with the AGN, \revv{and can be most naturally explained as the X-ray counterpart for the AGN}, further confirming the presence of an accreting black hole at the centre of NGC~3621. The low X-ray luminosity of the source can be used to place a lower limit on the black hole mass of $4\times10^3\,{\rm M_\odot}$ \citep{gliozzi09}.

\revv{However,} further investigation of the X-ray properties of this source is prevented by its low luminosity and the presence of the other, much brighter, X-ray sources in its immediate vicinity. X1's detections with \swift\ are unreliable and the upper limits obtained with both \swift\ and \xmm\ are too shallow to place any constraint on the source's long-term variability. \revv{Hence its association with a nuclear X-ray binary cannot be ruled out.}

\subsection{X2: the brightest source} 

X2 has been significantly detected in the three observations we presented. Interestingly, the spectral properties of the source changed over time: while in CXO1 and XMM1 the spectrum is consistent with an absorbed non-thermal emission, the spectrum observed in XMM2 is significantly softer and consistent with black-body-like emission from an accretion disk. 

\revv{Prompted by the fact that the statistical quality of the fit in XMM2 gets marginally improved by adding a second thermal component, we tried fitting the same two-thermal-component models to the previous epochs, obtaining satisfactory fits. While the temperatures of the soft and hard components are always in the $0.2-0.5$~keV, $1-3$~keV range, typical of ULXs \citep{stobbart06,sutton12,gurpide21}, the luminosity of the soft component drops by an order of magnitude in XMM2, with the hard components getting brighter by a factor of $\approx2.5$.}

\rev{While ULXs display a variety of spectra from hard ($\Gamma\approx1$ if modelled with a power law) to soft ($\Gamma\approx3$) with no evidence of bimodality \citep[see e.g.][]{feng11}, such} a change is reminiscent of the accretion states commonly observed in both persistent and transient Galactic X-ray disk-fed binaries, where a BH or NS system may be \rev{found} in a hard, soft, or intermediate state (see, e.g., \citealt{DeMarco2022}). In this scenario, X2 was consistent with being in the hard state in 2008 and 2017, and in the soft state\rev{, or, more likely, in a transition state in 2023.  Indeed, the ``canonical'' high/soft state seems to be rare in ULXs \citep{Soria2009}. If the parallel holds, in Galactic BH binaries the transition typically takes place at $\approx$30--50\% of the Eddington luminosity, suggesting an extremely heavy stellar or intermediate-mass BH.}
\revv{On the other hand, the possibility of reproducing the spectrum with two thermal components suggests that, similarly to other ULXs, X2 can be interpreted as a case of super-Eddington accretion on a stellar mass compact object.}

Unfortunately, the lack of features in the PDS of the source (most likely due to a limited signal-to-noise ratio) prevents us from \revv{confirming or ruling out either scenario} by employing information from the time variability domain. 

\subsection{X3: a new eclipsing ULX}\label{sec:discX3}

The detection of the egress of an eclipse in the light curve of X3 allows us to identify this source as one of the few \rev{known} eclipsing ULXs \citep[see Sect.~2.3 of][]{Fabrika2021}. We can set a lower limit on the eclipse duration of $\approx50$\,ks. Regarding the orbital period of the system, we can assume as a lower limit the duration of the whole observation $\approx90$\,ks. \rev{The presence of an eclipse, moreover, sets a lower limit on the inclination angle of the system $i\gtrsim75^\circ$, with respect to the line of sight.}

It is interesting to compare X3 with the \rev{two} eclipsing ULXs in M51: CXOM51\,J132940.0+471237 \rev{and} CXOM51\,J132939.5+471244 \citep[ULX--1 and ULX--2, respectively, in][]{Urquhart2016}.
\rev{CXOM51\,J132943.3+471135 and CXOM51\,J132946.1+471042
\citep[S1 and S2, respectively, in][]{Wang2018}, two other sources in M51 included among the ULXs showing eclipses by \cite{Fabrika2021}, have X-ray luminosities $L_\mathrm{X}\lesssim10^{39}$\,erg\,s$^{-1}$ \citep{Wang2018}. It is highly questionable whether they are indeed ULXs, therefore we will not consider them in our discussion.} 
In the case of ULX--1 \rev{and} ULX--2, \cite{Urquhart2016} could only set a lower limit of 70 \rev{and} 48\,ks, respectively, for the eclipse duration. The detection of an ingress and egress in two observations separated by 11 days allowed \cite{Urquhart2016} to constrain the orbital period of ULX--1 to $\approx6$\,d or $\approx12$\,d. 
\rev{Thanks to their constraints on the orbital period, }\cite{Urquhart2016} \rev{could infer} a mass of the companion star \rev{of ULX--1} in the range $M_\mathrm{co}\sim7-31{\rm M_\odot}$, \rev{typical of a high-mass X-ray binary (HMXB)}.

\rev{ULX--1 shows an evolution in the hardness ratio, with a softer emission during the eclipse. This is expected, since during an eclipse the central compact object (where the hard X-ray emission arises) is obscured by the companion. Unfortunately, in the case of X3, we could not derive a hardness ratio for the first 40~ks of the eclipse, due to the low number of collected source counts. We can expect, however, a similar evolution in the hardness ratio. Future detections of the eclipse in different observations could allow us to derive a stacked hardness ratio during and outside this phase and verify whether X3 is indeed softer during the eclipse.}
The presence of a dip at the end of the egress of X3 is probably due to inhomogeneous winds arising from the companion. A similar egress pattern is observed also in LMC X--4, a HMXB ($M_\mathrm{co}\simeq18$\,M$_\odot$) hosting a NS, often observed at super-Eddington luminosities \citep[see][and references therein]{Jain2024}. Given the similarity between these systems and X3, it is reasonable to assume that X3 is a HMXB whose companion mass is in a comparable range. 

\rev{The paucity of data regarding this source prevents us from a deeper discussion. The lower limit on the orbital period ($\lesssim90$\,ks) is shorter than any HMXB orbital period. At the same time, the time span between the two \xmm\ observations prevents us from deriving meaningful constraints. Finally, no optical counterpart is known for this source, therefore no radial velocity estimates (which could be used to infer the mass function) are available. Deeper observations are needed to further characterize this peculiar ULX.}

\subsection{X4: the (not-currently) beating ULX}

In the data from 2023, X4 was detected exhibiting a luminosity that aligns with the lower extremity of the spectrum observed towards the end of 2017 \citep{motta20}, only slightly below $10^{39}$ erg/s. In the 2017 data, this luminosity corresponds to emission phases occurring in the intervals between flares. The spectrum obtained in 2023 is also consistent with those obtained during the low-activity phases in 2017, in between flares. 

Unlike the observations from 2017, the most recent light curve analysis does not display any flaring activities. The Power Density Spectrum (PDS) lacks any indication of quasi-periodic modulations around the previously identified heartbeat period of approximately 3500 seconds (T $\approx$3500s), with a three-sigma (3$\sigma$) upper limit for the QPO rms amplitude at 0.28.

When comparing X4 to the archetype of heartbeat sources, GRS 1915+105 \citep{Belloni97,Belloni00}, the recently observed behaviour is not unexpected. In GRS 1915+105, similar accretion rates or luminosity levels sometimes coincide with heartbeat events, suggesting that the instability at the base of the flaring is not solely triggered by specific accretion rates. Other factors, such as certain disc properties (e.g. opacity or ionization state), may play a role in triggering the relevant instability. If the flaring observed in X4 in 2017 shares the same nature as the heartbeats in GRS 1915+105, it could represent an erratic phenomenon that might reoccur in the future at the same or at a different luminosity level. 

\subsection{X5: peek-a-boo}

The source labelled X5 is convincingly detected only in the first \cxo\ observation, showing an X-ray spectrum that can be well-modelled by a power law with photon index $\Gamma=1.6$ and a luminosity of about $2.3\times10^{39}$ erg/s, placing it amongst the ULXs. For reasons similar to the one discussed for X1, the \swift\ detections are unreliable and the only significant upper limits are the ones obtained with \xmm\ which indicates that the source's flux dropped by a factor $>5$.

Based on the source X-ray spectrum and luminosity, \citet{gliozzi09} estimated its mass as $2\times10^{3-4}\, {\rm M_\odot}$, interpreting this source as an accreting IMBH. However, today's interpretation of ULXs strongly favours a scenario in which these sources are powered by super-Eddington accretion onto neutron stars (see e.g. \citealt{kaaret17} and \citealt{Fabrika2021} for a review on the topic). 
Indeed, although the bulk of the ULX population shows steady levels of accretion, a subset exhibits strong variations in their flux, as wide as one order of magnitude, which could be a potential tell-tale of ULX pulsars (e.g. \citealt{earnshaw18,song20}). 
\rev{Another viable explanation for the behaviour of this source is an outburst by an otherwise normal transient X-ray binary, peaking in the ULX regime, as observed in a couple of objects in M31 \citep{middleton12,middleton13,esposito13}.}

\subsection{X6: the faintest sister}

The source labelled X6 is the faintest of the population of ULXs in the central region of NGC~3621. Its location fell out of the field of view in the first \cxo\ observation and in between two detector chips in the EPIC-pn on the first \xmm\ visit. Although its luminosity lies below the $10^{39}$~erg/s in both \xmm\ observations, it passes the threshold in at least one of the \swift/XRT visits.

Over the different epochs of the \xmm\ observations, the X-ray flux of X6 remains quite stable, varying only by a factor of about 0.7. No significant change is observed in the X-ray spectral shape of the source: the spectrum in the two available epochs can be satisfactorily reproduced by a power law with a common slope of $\Gamma\approx1.6$ \revv{or with a two thermal component model with temperatures compatible with the population of known ULXs.}

X6 does not exhibit any clear short-term variability in any of the \xmm\ observations, but the source is too faint to put significant constraints on the pulsed fraction of the signal.

Finally, as the other ULXs in NGC~3621, X6 has no clear or bright optical counterpart as too many faint sources crowd the circle error of the \xmm\ detection.

\section{Summary}\label{sec:summ}
In this paper, we reported the decade-long behaviour of the population of X-ray bright sources hosted in the nuclear region of NCG~3621.
We analyzed the long- and short-term variability as well as the spectral evolution of six sources, exploiting archival \cxo\ data, an extensive \swift\ monitoring campaign as well as recently acquired \xmm\ data.
The main results of our analysis are the discovery of a new eclipsing binary ULX (X3) and that the "heart-beating" ULX (X4) is currently in an intermediate flux level but it is not showing its characteristic pulsation.
One source (X2) exhibits, across different \xmm\ observations, changes in its spectral shape with no significant variation in its flux levels, a behaviour common in HMXBs \revv{and ULXs}.
Two sources (X1 and X5) are not detected in any \xmm\ or \swift\ visit, and the only available data comes from archival \cxo\ observation.
The picture is complete by the source labelled X6, which, although the faintest of the studied population, still falls in the ULX regime.

\begin{acknowledgements}
We thank the anonymous referee for their comments which improved the quality of the paper. M.I. is supported by the AASS Ph.D. joint research programme between the University of Rome ``Sapienza" and the University of Rome "Tor Vergata", with the collaboration of the National Institute of Astrophysics (INAF). PE, GLI, and AT acknowledge financial support from the Italian Ministry for University and Research, through the grants 2022Y2T94C (SEAWIND) and from INAF through LG 2023 BLOSSOM. GLI acknowledges financial support from INAF through grant ``INAF-Astronomy Fellowships in Italy 2022 - (GOG). This work was partially supported by the ASI--INAF program I/004/11/5.''
\end{acknowledgements}  

\bibliographystyle{aa} 
\bibliography{biblio} 

\appendix

\section{Particle Swarm Optimization}
Particle Swarm Optimization (PSO) is a meta-heuristic optimization evolutionary algorithm, introduced in \cite{Kennedy}, that can search for the maximum value of a function $f: \mathbb R^n \to \mathbb R$ whose analytical expression is unknown. In this work, there are $n = 4$ independent variables, corresponding to 4 orbital parameters (i.e., $T_\textup{orb}$, $T_\textup{asc}$, $a_\textup{X}\sin i$, and $\dot P/P$). Let $C_\textup{s}$ be the lightcurve corrected according to a set of orbital parameters $s$ and $P_\textup{s}$ the corresponding power density spectrum (computed using \texttt{stingray} \citealt{Huppenkothen2019}). Then, the dependent variable of $f$ is $\max(P_\textup{s})$ (i.e., the highest power peak in $P_\textup{s}$). Unlike grid search, PSO does not require pre-defining a search grid, which would require making assumptions on the grid step sizes.

Multiple searches are performed on each observation with different PSO hyperparameter values for $c_1$ (nostalgia), $c_2$ (envy), and $w$ (inertia). In particular, $c_1 \in \{0.5, 1.0, 1.5\}$, $c_2 \in \{0.5, 1.0, 1.5\}$, and $w \in \{0.5, 0.75, 1\}$, for a total of 27 combinations per observation. Different searches consider three different scenarios. If $c_1 > c_2$, the algorithm privileges the exploration of diverse candidate sets of orbital parameters; if $c_1 < c_2$, the algorithm tends to converge towards local maxima more prematurely, and if $c_1 \approx c_2$, the algorithm balances exploration and convergence towards a maximum solution.

Different results obtained in different searches suggest that (1) the highest powers are not statistically significant ($< 3\sigma$), and (2) the optimal orbital parameters combination and frequency vary depending on the search (and not on the source) while having similar powers. Combined with the $\dot P$ analysis, this approach suggests that the signal is either absent or indistinguishable from noise.

\section{X-ray observations}
The journal of the X-ray observations is given in Tab. B.1\footnote{ The Table is available at \url{https://doi.org/10.5281/zenodo.12582646}.}

\section{Light curves and PDSs of \xmm\ observations.}

\rev{In this section\footnote{All plots are available at \url{https://doi.org/10.5281/zenodo.12582691}.} we report the remaining light curves and PDSs not shown in the main text.} 

\end{document}